\documentclass[aip,jcp,reprint,amsmath,amssymb]{revtex4-1}
\usepackage{graphicx}
\usepackage{dcolumn}
\usepackage{bm}
\usepackage[utf8]{inputenc}
\usepackage[T1]{fontenc}
\usepackage{mathptmx}
\usepackage{etoolbox,float}
\usepackage{booktabs}

\makeatletter
\def\@email#1#2{%
 \endgroup
 \patchcmd{\titleblock@produce}
  {\frontmatter@RRAPformat}
  {\frontmatter@RRAPformat{\produce@RRAP{*#1\href{mailto:#2}{#2}}}\frontmatter@RRAPformat}
  {}{}
}%
\makeatother

\begin{document}
\preprint{}

\title{Calculations of pathways of precise P incorporation into chlorinated Si(100) surface}

\author{T. V. Pavlova}
 \email{pavlova@kapella.gpi.ru}
\affiliation{Prokhorov General Physics Institute of the Russian Academy of Sciences, Vavilov str. 38, 119991 Moscow, Russia}
\affiliation{HSE University, Myasnitskaya str. 20, 101000 Moscow, Russia}

\date{\today}

\begin{abstract}
The precise incorporation of a phosphorus atom into a silicon surface is essential for the fabrication of nanoelectronic devices in which the active area is formed from single impurities. The most accurate approach employs scanning tunneling microscopy (STM) lithography, which may be done with atomic precision. However, the accuracy decreases when phosphorus is incorporated into the surface because P substitutes one of two neighboring Si atoms with equal probability. Here, the P-Si exchange mechanism was studied theoretically on a chlorinated Si(100) surface with an asymmetric configuration of Cl vacancies surrounding the P atom. Density functional theory was used to estimate the activation barriers and exchange rates between a P atom and neighboring Si atoms on a Si(100)-2$\times$1-Cl surface with three Cl vacancies. The calculation of various P-Si exchange pathways revealed that phosphorus has a higher probability of substituting one Si atom than the others due to the asymmetric configuration of Cl vacancies. Based on the theoretical study of the P-Si exchange mechanism and experimental results from previous works, a scheme for controlled P incorporation into the silicon surface without uncertainty is proposed.

\end{abstract}

\maketitle

\section{Introduction}

Precise incorporation of impurities into silicon allows the creation of new types of electronic devices. In particular, phosphorus placed with atomic precision is highly desirable for quantum computing in silicon, which utilizes the nuclear and electron spins of donor atoms \cite{1998Kane}. Currently, the most precise doping is used to fabricate quantum systems, including elements of quantum computers \cite{2019He}, single-atom nanoelectronic devices \cite{2012Fuechsle}, and artificial lattices in semiconductors \cite{2022Kiczynski, 2022Wang, 2020Le}.

The method of silicon precise doping with phosphorus exploits the reaction between phosphine and a Si(100) surface covered with a hydrogen monolayer \cite{2003Schofield}. To generate active sites for PH$_3$ adsorption, single hydrogen atoms are desorbed using STM lithography \cite{2017Moller, 2018Achal, 2018Randall}. The minimum window in a hydrogen monolayer resist for PH$_3$ dissociative adsorption on PH$_2 +$H consists of a Si dimer \cite{2016Warschkow}, which is formed by removing two neighboring H atoms. The most precise method of P incorporation involves the removal of H atoms by an STM tip from both the Si(100)-2$\times$1-H surface and the adsorbed phosphine molecule, resulting in full PH$_2$ dissociation \cite{2022Wyrick}. The removal of H atoms from the silicon surface \cite{2017Moller, 2018Achal} and from an adsorbed phosphine molecule \cite{2022Wyrick, 2016Liu} is possible with atomic precision.

In the following step of the precision doping method, the phosphorus atom exchanges places with a surface silicon atom, becoming a substitutional impurity \cite{2003Schofield}. To initiate this activation process, the surface must be heated. Before annealing, the P atom locates in the most stable adsorption position between two Si atoms of adjacent dimers in one dimer row (end-bridge position) and upon annealing P exchanges place with one of two equivalent adjacent Si atoms \cite{2009Bennett}. Thus, the exchange mechanism has uncertainty in P incorporation of two adjacent lattice sites.

To predetermine the phosphorus incorporation process, this study proposes to make the Si surface atoms around the P atom non-equivalent. To make the Si atoms non-equivalent, the adsorbate atoms can be arranged in an asymmetric configuration. Chlorine was chosen as the adsorbate and PBr$_3$ as the phosphorus-containing molecule, since they had previously been shown to be promising for the reaction of phosphorus incorporation into silicon surface \cite{2022Pavlova, 2023Shevlyuga, 2024Pavlova}. Here, the exchange mechanism of phosphorus with one of the nearest Si atoms on the Si(100)-2$\times$1-Cl surface with three Cl vacancies is considered theoretically. The calculation of activation barriers for different P-Si exchange pathways revealed that one of the pathways has the lowest barrier, and the reaction rate along this pathway is several orders of magnitude higher than the rates of the P exchange with other Si atoms. As a result, controlled exchange of phosphorus with a specific rather than arbitrary neighboring Si atom can be possible. Based on the calculations and experimental studies from previous works \cite{2022Pavlova, 2024Pavlova}, a scheme for exact site P incorporation into Si(100) is proposed.

\section{Computational details}

The spin-polarized DFT calculations were performed with the Vienna \textit{ab initio} simulation package (VASP) \cite{1993Kresse, 1996Kresse} utilizing the projector augmented wave method  \cite{1999Kresse, 1994Blochl} and the Perdew-Burke-Ernzerhof (PBE) generalized gradient approximation \cite{1996Perdew}. The slabs for the Si(100)-2$\times$1 surface were chosen to be periodic 4$\times$4 supercells with eight atomic Si layers and a 15\,{\AA} vacuum space. A Si(100)-2$\times$1-Cl structure was formed by the placement of Cl atoms on the top Si surface, while hydrogen atoms covered the lower Si surface. The bottom three Si layers and H atoms were kept fixed while the rest atoms were allowed to relax. The slabs were relaxed until the residual force fell below 0.01\,eV/\,{\AA}. A $\Gamma$-centered 3$\times$3$\times$1 k-point mesh and the plane-wave cutoff energy of 350\,eV were used in calculations. The activation barrier calculations were performed utilizing the nudged elastic band (NEB) method \cite{1998NEB}, and the convergence criteria for residual forces was decreased to 0.03\,eV/\,{\AA}. Calculations for three main structures (P1, P2, and Si1) and two paths (D and T4) were also carried out for the 18-layer slab. The 8- and 18-layer slabs have identical structures and pathways, with an estimated error of 5 meV. The reaction rate was estimated according to the Arrhenius equation as it was done in previous work \cite{2023Shevlyuga}. Note that the activation barrier, which has a computational error, is included in the exponent in the Arrhenius equation. Therefore, the reaction rates are estimated rather than exact values.

\section{Results}

\subsection{Adsorption positions of phosphorus}

\begin{figure}[h]
 \includegraphics[width=\linewidth]{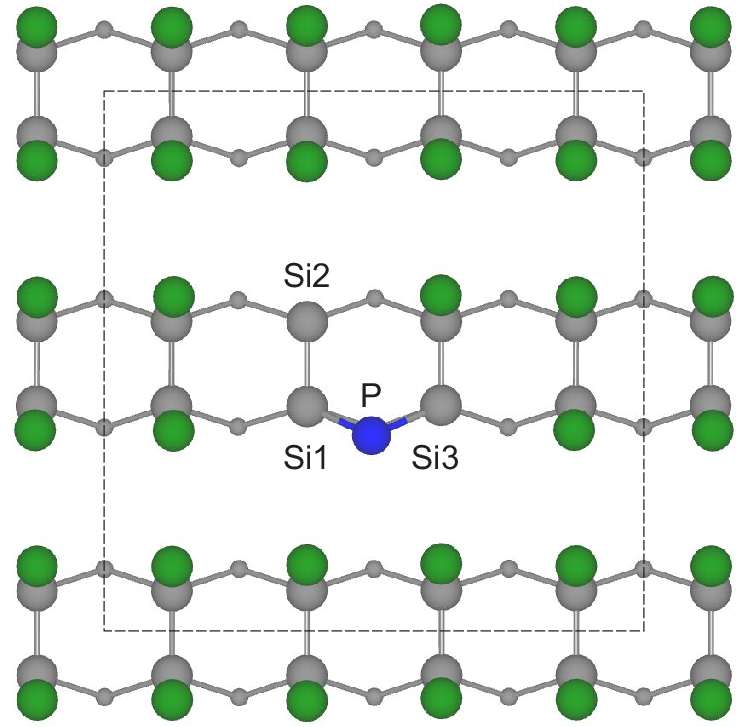}
\caption{\label{fig1} Model of the Si(100)-2$\times$1-Cl surface with three Cl atoms removed and an adsorbed P atom, shown as an example in the end-bridge position. The Si atoms in the top layer are represented by big gray circles, Si in the second layer by small gray circles, Cl by green, and P by blue. The dotted lines indicate the supercell for which the calculations were done.}
\end{figure}

Figure~\ref{fig1} shows a Si(100)-2$\times$1 surface with an adsorbed P atom and a chlorine monolayer containing three Cl vacancies. The three silicon atoms (Si1, Si2, and Si3) next to the P atom have different Cl surroundings. The asymmetric configuration of Cl vacancies is designed to create unequal P exchange pathways with nearby Si atoms.

\begin{figure}[h]
 \includegraphics[width=\linewidth]{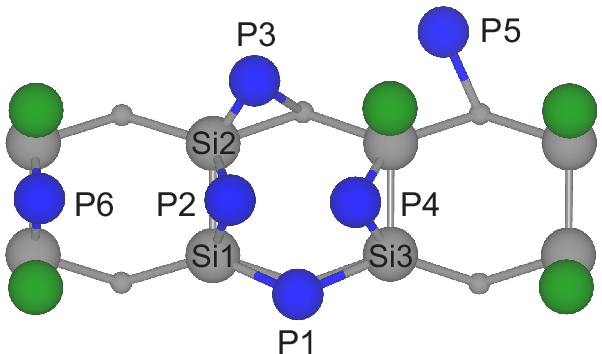}
\caption{\label{fig2} Different adsorption positions of the P atom on the Si(100)-2$\times$1-Cl surface with three Cl vacancies. The Si atoms in the top layer are shown as large gray circles, Si in the second layer as small gray circles, Cl in green, and P in blue.}
\end{figure}

To find the preferred adsorption site of phosphorus on such a surface, the adsorption energies for six different positions of the P atom were calculated (Fig.~\ref{fig2}). In position P1, phosphorus bonds with the Si1 and Si3 atoms and the Si atom of the previous layer above which it is located (end-bridge position). In position P2, the phosphorus atom forms bonds with the Si1 and Si2 atoms of the silicon dimer (bridge position). In the bridge position, P has two symmetrical adsorption sites, slightly shifted to the right or left relative to the Si dimer, with a very low activation barrier for the transition between these positions \cite{2009Bennett}.

In the next two positions, one of the Si atoms adjacent to the P atom has a bond with chlorine. At position P3, the P atom was placed at the end-bridge position symmetrical to P1, but was displaced from it as a result of geometry optimization. In contrast to P1, in P3 phosphorus has formed bonds with only two of the three Si atoms that do not have bonds with Cl.
In P4, the P atom was placed in the bridge position between the Si dimer with one Cl atom, and it was moved away from the center of the dimer during the optimization process.

In the remaining two positions, two Si atoms located close to the P atom are bonded to chlorine. In this case, the P atom does not have a local minimum at the end-bridge position and, during geometry optimization, it shifted into the groove between the dimer rows (position P5). In P5, phosphorus has two bonds with the Si atoms of the previous layer and with the Si atom of the third layer (only one bond with the Si atom of the second layer is shown in Fig.~\ref{fig2} since the next dimer row and the atoms of the third Si layer are not displayed). At position P6, the phosphorus atom is located between a silicon dimer containing two chlorine atoms.

Table~\ref{table1} shows the difference in P adsorption energies at positions P1--P6. The end-bridge position P1 is the most favorable, followed by the bridge position P2, which is consistent with the results of previous work on a clean Si(100) surface \cite{2009Bennett}. These positions were found to be less stable when one of the adjacent Si atoms was bonded to a Cl atom (P3, P4), and even less stable when both nearest Si atoms were bonded to chlorine (P5, P6). Note that the P6 position is very unfavorable because the P atom inserted into the chlorinated Si dimer pushes the Si and Cl atoms of the dimer apart, which leads to repulsion between the Cl atoms of neighboring dimer rows. This problem is caused by the small supercell, which contains only two dimer rows (Fig.~\ref{fig1}). To estimate the adsorption energy of phosphorus at the P6 position without the impact of the repulsive interaction between chlorine, two Cl atoms on the neighboring dimer row were transferred to the Si1 and Si2 atoms. In this case, the relative adsorption energy reduced to 1.40 eV.

\begin{table}[h]
\begin{center}

    \begin{tabular}{|c|c|}   \hline
    P adsorption site & Relative adsorption energy, eV \\ \hline \hline
    P1 & 0.00\,eV    \\ \hline
    P2 & 0.49\,eV    \\ \hline
    P3 & 1.29\,eV    \\ \hline
    P4 & 1.21\,eV    \\ \hline
    P5 & 2.51\,eV    \\ \hline
    P6 & 2.11\,eV (1.40\,eV)   \\ \hline

\end{tabular}
\caption{Relative adsorption energies of phosphorus in the positions shown in Fig.~\ref{fig2} on the Si(100)-2$\times$1-Cl surface with three Cl vacancies. The adsorption energy is given relative to the most favorable position P1. For position P6, the relative adsorption energy in brackets is given in the absence of repulsive interaction between chlorine atoms (when two Cl atoms of the adjacent dimer row with the Si dimer containing P move to the Si1 and Si2 atoms).}
    \label{table1}
\end{center}
\end{table}

\subsection{P-Si exchange pathways}

The P-Si exchange mechanism is considered starting from positions P1 and P2, which are significantly more stable than the others. To combine the exchange mechanisms from the two initial positions P1 and P2 into one energy diagram, the minimum energy path for P diffusion between these positions was calculated. The obtained activation barrier for P diffusion on the Si(100)-2$\times$1-Cl surface with three Cl vacancies is 1.21\,eV. This value is slightly higher compared to the activation barriers for P diffusion on a clean Si(100) surface reported in previous studies, 0.94\,eV \cite{2009Bennett} and 0.8\,eV \cite{1992Brocks}.

\begin{figure}[h]
 \includegraphics[width=\linewidth]{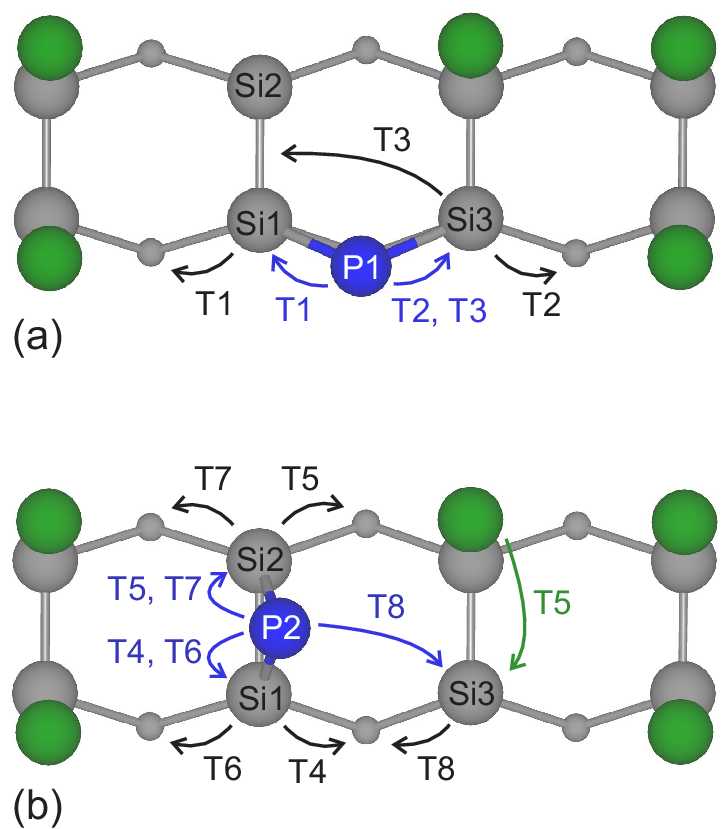}
\caption{\label{fig3} The P-Si exchange pathways with phosphorus located in the end-bridge (a) and bridge (b) adsorption positions. The Si atoms in the top layer are represented by big gray circles, Si in the second layer by small gray circles, Cl by green, and P by blue. The blue, black, and green arrows show the paths of the P, Si, and Cl atoms, respectively.}
\end{figure}

The P-Si exchange pathways are shown schematically in Fig.~\ref{fig3}, and their activation barriers are given in Table~\ref{table2}. In pathways T1 and T2, phosphorus from position P1 substitutes the adjacent Si atom, Si1 and Si3, respectively, and the substituted Si atom occupies the adjacent end-bridge position (Fig.~\ref{fig3}a). In the T3 pathway, the substituted silicon atom Si3 moves to the bridge position.

\begin{table}[h]
\begin{center}

    \begin{tabular}{|c|c|c|c|c|}   \hline
    Pathway & P position & Substituted Si & E$_{FS}$, eV& E$_a$, eV \\ \hline \hline
    T1 & P1 & Si1 & 1.82 & 2.23   \\ \hline
    T2 & P1 & Si3 & 1.43 & 1.97   \\ \hline
    T3 & P1 & Si3 & 0.18 & 2.59   \\ \hline
    T4 & P2 & Si1 & -0.01 & 1.16   \\ \hline
    T5 & P2 & Si2 & -0.01 & 1.42   \\ \hline
    T6 & P2 & Si1 & 1.33 & 1.44   \\ \hline
    T7 & P2 & Si2 & 1.30 & 1.42   \\ \hline
    T8 & P2 & Si3 & -0.02 & 2.04   \\ \hline

\end{tabular}
\caption{Activation barriers for the exchange pathways of a P atom with a Si atom from the initial positions P1 and P2. For each pathway T1--T8, the initial position of phosphorus, the substituted Si atom, the energy of the final state (E$_{FS}$), and the activation barrier (E$_a$) are given. E$_{FS}$ are calculated relative to the initial states energies (P1 or P2).}
    \label{table2}
\end{center}
\end{table}

In the T4--T8 pathways, the initial position of phosphorus is P2, and the final position of the substituted Si atom is the end-bridge position (Fig.~\ref{fig3}b). The T4 pathway has the lowest activation energy of 1.16 eV. Note that the same P-Si exchange pathway from the bridge position on a clean Si(100) surface has the lowest barrier of 1.06 eV \cite{2009Bennett}. In the T5 pathway, the final position of Si2 is less favorable than in the T4 pathway because the neighboring Si atom bonded to the Cl atom has four bonds, making the creation of a new bond unfavorable. To make end-bridge position of Si2 more advantageous, the Cl atom was transferred to the Si3 atom. For the same reason, the final states of pathways T6 and T7 are less favorable. The final state of T8 is as favorable as in T4, but the barrier is higher than in pathways T4--T7 (Table~\ref{table2}).

\begin{figure*}[t]
 \includegraphics[width=\linewidth]{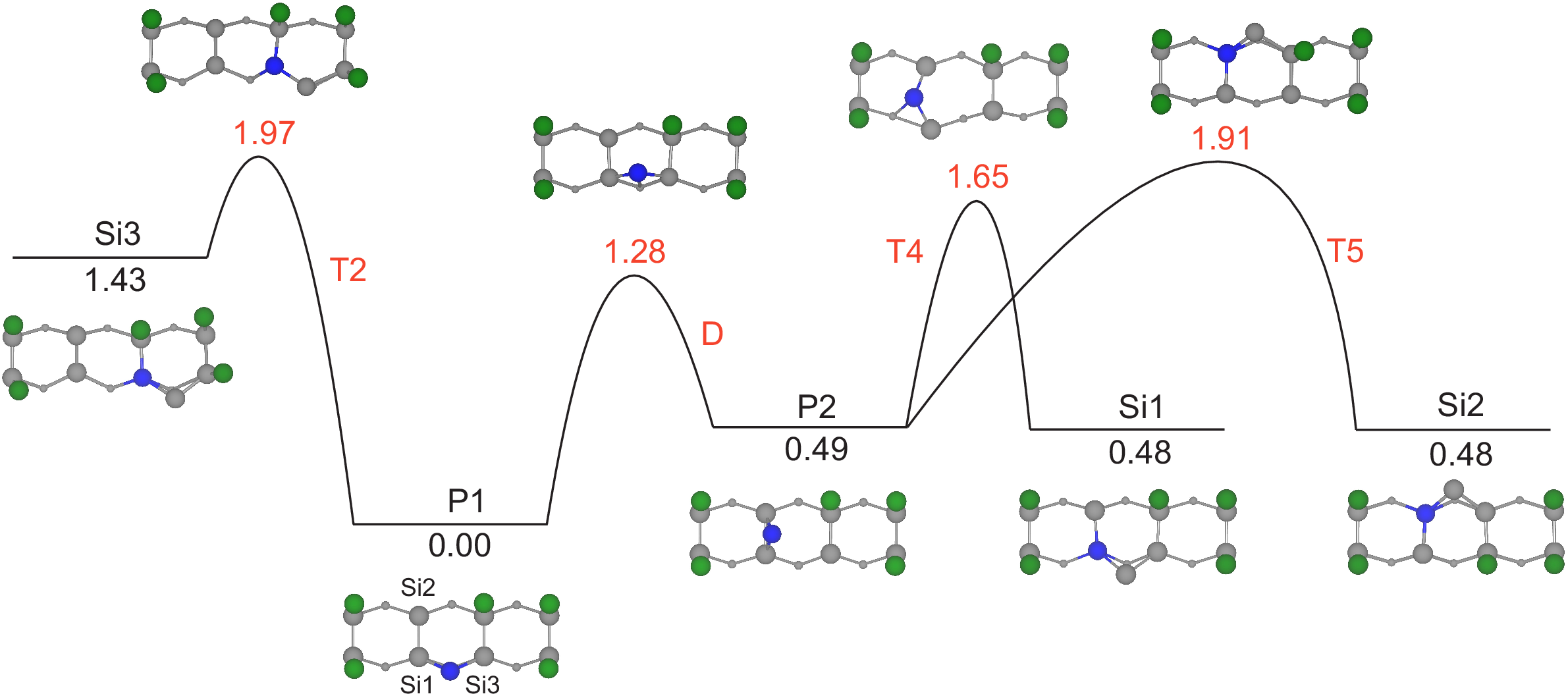}
\caption{\label{fig4} Energy barrier diagram of the P-Si exchange. Only the most favorable pathways for P exchange with Si1, Si2, and Si3 are shown. All energies are given in electronvolts relative to the energy of the initial position P1. The energies of initial and final states are shown in black, and activation barriers in red. The structures of initial and final states are shown at the bottom, and the structures of transition states in the upper part of the diagram. The Si atoms in the top layer are shown as large gray circles, Si in the second layer as small gray circles, Cl in green, and P in blue.
}
\end{figure*}

Figure~\ref{fig4} shows the energy barrier diagram of the P-Si exchange, displaying only the most favorable exchange pathways for each of the silicon atoms Si1, Si2, and Si3. The phosphorous diffusion (D pathway) connects the P-Si exchange pathways with initial positions P1 and P2. Phosphorus at position P1 has the lowest barrier when substituting Si1 (paths D and then T4), followed by Si2 (paths D and then T5), and the highest barrier when substituting Si3 (T2). The sequence remains unchanged for the initial positions P2; the barrier is lowest when P substitutes Si1 (T4), then Si2 (T5), and then Si3 (paths D and then T3). Thus, the activation barriers are different when phosphorus substitutes Si1, Si2, and Si3.

\section{Discussions}

\begin{figure*}[t]
 \includegraphics[width=\linewidth]{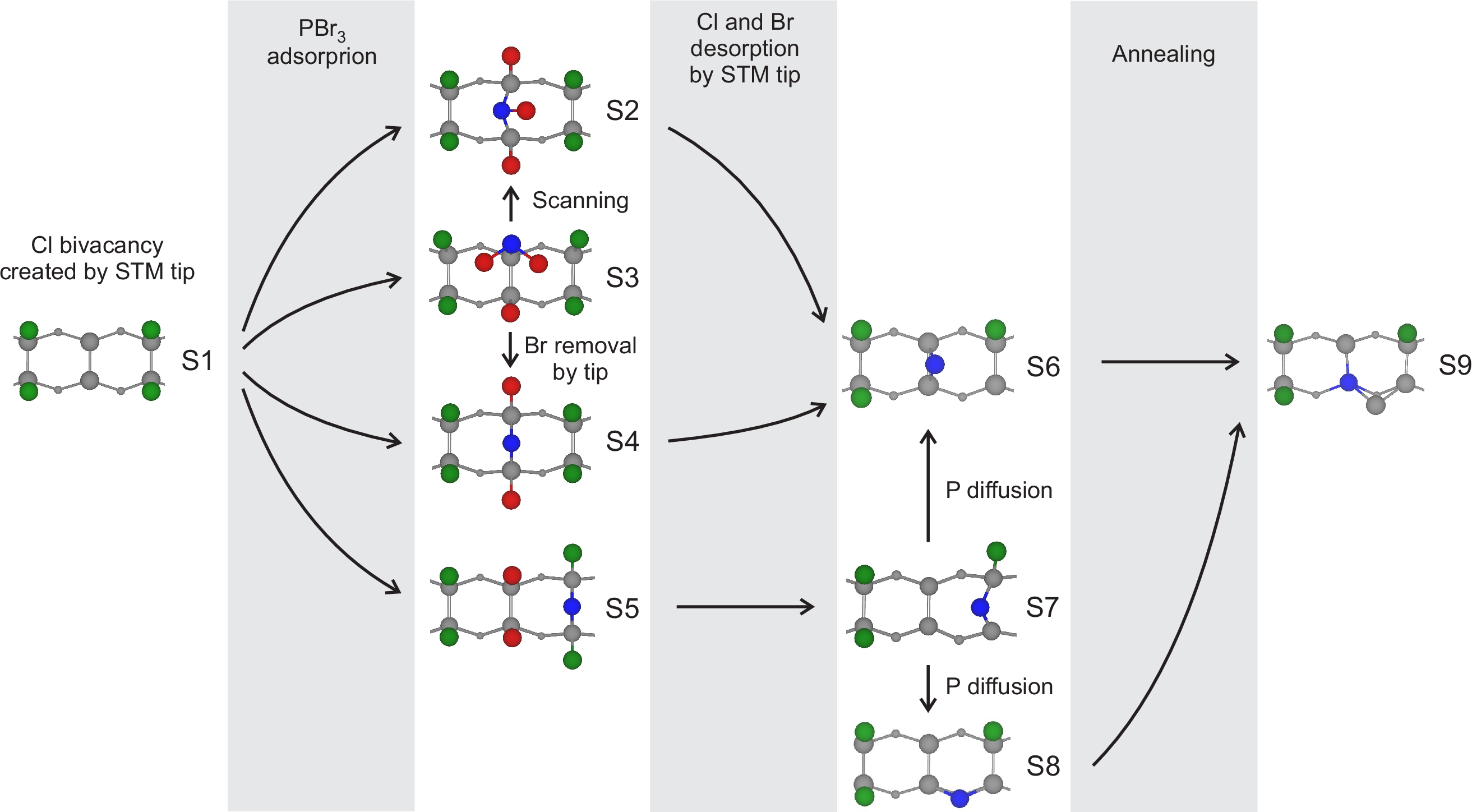}
\caption{\label{fig5} Proposal for atomically precise substitution of the Si atom by phosphorus. An STM tip is utilized to create a Cl bivacancy on Si(100)-2$\times$1-Cl (structure S1), into which a PBr$_3$ molecule is adsorbed. The dissociated molecule in the Cl bivacancy forms structures S2--S5 with phosphorus \cite{2024Pavlova}. Structure S3 can be converted to S2 by scanning or to S4 by removing a Br atom from PBr$_2$ with a tip. The removal of one Cl and all Br atoms transforms S2 and S4 to S6. Structure S5 converts to S7, from which the P atom can diffuse to more favorable positions in structures S6 or S8. After annealing, S6 and S8 convert to S9. The Si atoms in the top layer are shown as large gray circles, Si in the second layer as small gray circles, Cl in green, Br in red, and P in blue.}
\end{figure*}

Based on the obtained result on the influence of the adsorbate environment on the P-Si exchange and on previous experimental studies, a scheme of atomically precise P incorporation into Si(100) is proposed below (Fig.~\ref{fig5}). The novel approach is to create an asymmetric structure from adsorbate atoms before the sample annealing, while the idea of tip-assisted incorporation was adapted from Ref.~\cite{2022Wyrick}.

Initially, a Cl bivacancy is created on the Si dimer on the Si(100)-2$\times$1-Cl surface (structure S1 in Fig.~\ref{fig5}). The possibility of removing single halogen (Cl, Br) atoms from the Si(100)-2$\times$1-Cl (-Br) surface in STM has already been demonstrated previously \cite{2022Pavlova}. Then, the phosphorus tribomide is adsorbed to the surface. PBr$_3$ is considered for use as a phosphorus-containing molecule since a Cl monolayer most effectively protects the Si(100) surface from the undesirable incorporation of halogenated molecules \cite{2021Pavlova}. Phosphorus does not bond to silicon under a Cl monolayer or in single Cl vacancies, which either remain empty or are filled with Br if the vacancy was positively charged \cite{2023JCPPavlova}.

PBr$_3$ adsorption resulted in S2--S5 structures with phosphorus (Fig.~\ref{fig5}) \cite{2024Pavlova}. The PBr$_3$ molecule dissociates in the Cl bivacancy so that the P atom occupies a position above the Si atom (S3) or bridge position (S2, S4, S5). For further incorporation the P atom should be positioned at the bridge site, hence structure S3 needs to be transformed to structure S2 or S4. Structure S2 is more stable than S3 by 0.81 eV and such a transformation from S3 to S2 was observed experimentally during scanning \cite{2024Pavlova}. Also, according to our calculations, structure S3 can be transformed into S4 by removing one Br atom from the PBr$_2$ fragment by an STM tip. The PBr fragment that remains after Br removal moves to the bridge position, which is 0.14 eV more favorable and breaks the Si-Si dimer bond with a barrier of 0.48 eV. Then, the Br atom from the PBr fragment moves to the Si atom, overcoming the barrier of 0.49 eV. The final state is 1.12 eV more favorable than the initial one, and the barriers are low enough for the process to occur at room temperature. Structure S3 is therefore can be transformed to S2 or S4.

In the next step, the STM tip is used to remove one Cl and all of the Br atoms. STM desorption of H atoms from the PH$_3$ molecule by the tip was originally realized on a clean Si(100) surface \cite{2016Liu}. This idea was further developed for tip-assisted incorporation of phosphorus by dissociation of PH$_2$ in H vacancies on the hydrogenated Si(100) surface \cite{2022Wyrick}. In this case, hydrogen atoms were removed both from the molecular fragment PH$_2$ and from the surface near PH$_2$. In the current proposal, in addition to all Br atoms, the Cl atom is removed from the adjacent dimer on the same side of the dimer row on which the P-Si exchange is desired. After removal of all Br atoms, the P atom in structures S2 and S4 remains in a bridge position (structure S6).

In the S5 structure, two Br atoms are attached to the Si dimer with a Cl bivacancy, while phosphorus is located on the neighboring dimer. After removal of both Br atoms and one Cl atom, phosphorus remains in the bridge position (S7). Phosphorus can diffuse to a 1.21 eV more favorable end-bridge position with a 0.38 eV barrier (S8) or to a bridge position on a dimer without Cl with an energy gain of 0.71 eV with a barrier of 0.49 eV (S6). Both of these processes have a final state that is energetically more favorable than the initial state, and their activation energies are low enough to occur at room temperature. Therefore, structure S7 is transformed either to S6 or S8 (which is more likely).

At the last step of the P incorporation, the surface is annealed. When the surface is annealed, the P atom from both S6 and S8 exchange with the Si1 atom (structure S9), according to Fig.~\ref{fig4}. Note that even if the P atom is adsorbed on the neighboring dimer (S5), it can substitute the desired Si atom of the dimer with a Cl bivacancy if the correct halogen atoms are removed with a tip.

\begin{figure}[h!]
 \includegraphics[width=\linewidth]{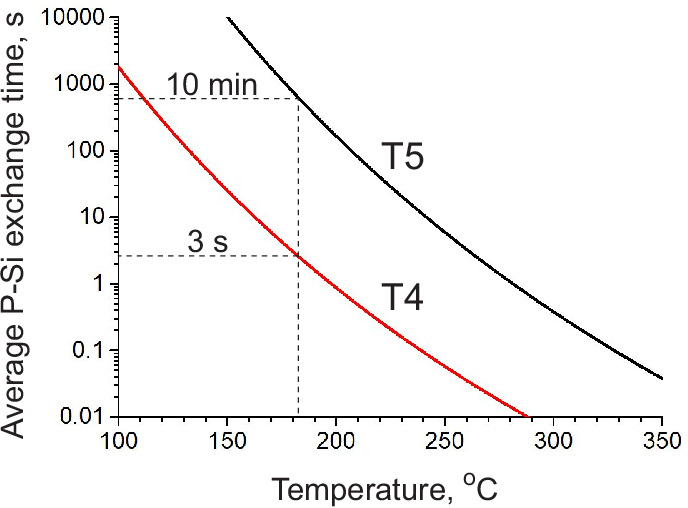}
\caption{\label{fig6} Comparison of the estimated average time of P-Si exchange for two pathways with minimal activation energy, T4 and T5. It is assumed that phosphorus was initially in the bridge position (P2). The activation barriers are 1.16 eV and 1.42 eV for T4 and T5, respectively (Table~\ref{table2}).}
\end{figure}

The temperature required for the P-Si exchange can be estimated using the Arrhenius equation by analogy with Ref.~\cite{2012Cui}. Figure~\ref{fig6} shows the dependence of the average P-Si exchange time on temperature for two pathways with minimal activation energy, T4 and T5, starting from the initial bridge position (P2). According to calculations, phosphorus in bridge position can substitute Si at temperatures about 180$^{\circ}$C. Note that the P-Si exchange from the end-bridge position (P1) requires a higher temperature, since the overall activation barrier is higher. The time of reaction for the two pathways differ by more than two orders of magnitude, 3 seconds for T4 and 10 minutes for T5, allowing the P-Si exchange to occur via the preferred T4 pathway.

\section{Conclusions}

A method for controlled P incorporation into the Si(100) surface is proposed. The key idea of this method is to create an asymmetric structure of Cl vacancies around the P atom using an STM tip. In the presence of three Cl vacancies surrounding an adsorbed P atom, incorporation occurs via a favored pathway of P exchange with only one Si atom. The implementation of this method should allow P atoms to be positioned on the surface with absolute precision. The main challenge for the experimental realization of this method is to improve the precision of single halogen atom extraction by an STM tip. Further, to create electronic devices, a Si epitaxial layer must be carefully formed so that the P atoms remain in their precise positions. Although this method has been proposed for halogens, it may be applicable to various adsorbates and molecules (for example, hydrogen monolayer and PH$_3$ molecule).

\section{Acknowledgments}
This research was supported in part through computational resources of HPC facilities at HSE University.

\section{Author declaration}
The author has no conflicts to disclose.

\section{Data availability}
The data that support the findings of this study are available from the corresponding author upon reasonable request.

\bibliography{Pavlova_JCP_rev}

\begin{thebibliography}{27}%
\makeatletter
\providecommand \@ifxundefined [1]{%
 \@ifx{#1\undefined}
}%
\providecommand \@ifnum [1]{%
 \ifnum #1\expandafter \@firstoftwo
 \else \expandafter \@secondoftwo
 \fi
}%
\providecommand \@ifx [1]{%
 \ifx #1\expandafter \@firstoftwo
 \else \expandafter \@secondoftwo
 \fi
}%
\providecommand \natexlab [1]{#1}%
\providecommand \enquote  [1]{``#1''}%
\providecommand \bibnamefont  [1]{#1}%
\providecommand \bibfnamefont [1]{#1}%
\providecommand \citenamefont [1]{#1}%
\providecommand \href@noop [0]{\@secondoftwo}%
\providecommand \href [0]{\begingroup \@sanitize@url \@href}%
\providecommand \@href[1]{\@@startlink{#1}\@@href}%
\providecommand \@@href[1]{\endgroup#1\@@endlink}%
\providecommand \@sanitize@url [0]{\catcode `\\12\catcode `\$12\catcode
  `\&12\catcode `\#12\catcode `\^12\catcode `\_12\catcode `\%12\relax}%
\providecommand \@@startlink[1]{}%
\providecommand \@@endlink[0]{}%
\providecommand \url  [0]{\begingroup\@sanitize@url \@url }%
\providecommand \@url [1]{\endgroup\@href {#1}{\urlprefix }}%
\providecommand \urlprefix  [0]{URL }%
\providecommand \Eprint [0]{\href }%
\providecommand \doibase [0]{http://dx.doi.org/}%
\providecommand \selectlanguage [0]{\@gobble}%
\providecommand \bibinfo  [0]{\@secondoftwo}%
\providecommand \bibfield  [0]{\@secondoftwo}%
\providecommand \translation [1]{[#1]}%
\providecommand \BibitemOpen [0]{}%
\providecommand \bibitemStop [0]{}%
\providecommand \bibitemNoStop [0]{.\EOS\space}%
\providecommand \EOS [0]{\spacefactor3000\relax}%
\providecommand \BibitemShut  [1]{\csname bibitem#1\endcsname}%
\let\auto@bib@innerbib\@empty
\bibitem [{\citenamefont {Kane}(1998)}]{1998Kane}%
  \BibitemOpen
  \bibfield  {author} {\bibinfo {author} {\bibfnamefont {B.~E.}\ \bibnamefont
  {Kane}},\ }\href {\doibase 10.1038/30156} {\bibfield  {journal} {\bibinfo
  {journal} {Nature}\ }\textbf {\bibinfo {volume} {393}},\ \bibinfo {pages}
  {133} (\bibinfo {year} {1998})}\BibitemShut {NoStop}%
\bibitem [{\citenamefont {He}\ \emph {et~al.}(2019)\citenamefont {He},
  \citenamefont {Gorman}, \citenamefont {Keith}, \citenamefont {Kranz},
  \citenamefont {Keizer},\ and\ \citenamefont {Simmons}}]{2019He}%
  \BibitemOpen
  \bibfield  {author} {\bibinfo {author} {\bibfnamefont {Y.}~\bibnamefont
  {He}}, \bibinfo {author} {\bibfnamefont {S.~K.}\ \bibnamefont {Gorman}},
  \bibinfo {author} {\bibfnamefont {D.}~\bibnamefont {Keith}}, \bibinfo
  {author} {\bibfnamefont {L.}~\bibnamefont {Kranz}}, \bibinfo {author}
  {\bibfnamefont {J.~G.}\ \bibnamefont {Keizer}}, \ and\ \bibinfo {author}
  {\bibfnamefont {M.~Y.}\ \bibnamefont {Simmons}},\ }\href {\doibase
  10.1038/s41586-019-1381-2} {\bibfield  {journal} {\bibinfo  {journal}
  {Nature}\ }\textbf {\bibinfo {volume} {571}},\ \bibinfo {pages} {371}
  (\bibinfo {year} {2019})}\BibitemShut {NoStop}%
\bibitem [{\citenamefont {Fuechsle}\ \emph {et~al.}(2012)\citenamefont
  {Fuechsle}, \citenamefont {Miwa}, \citenamefont {Mahapatra}, \citenamefont
  {Ryu}, \citenamefont {Lee}, \citenamefont {Warschkow}, \citenamefont
  {C~L~Hollenberg}, \citenamefont {Klimeck},\ and\ \citenamefont
  {Simmons}}]{2012Fuechsle}%
  \BibitemOpen
  \bibfield  {author} {\bibinfo {author} {\bibfnamefont {M.}~\bibnamefont
  {Fuechsle}}, \bibinfo {author} {\bibfnamefont {J.}~\bibnamefont {Miwa}},
  \bibinfo {author} {\bibfnamefont {S.}~\bibnamefont {Mahapatra}}, \bibinfo
  {author} {\bibfnamefont {H.}~\bibnamefont {Ryu}}, \bibinfo {author}
  {\bibfnamefont {S.}~\bibnamefont {Lee}}, \bibinfo {author} {\bibfnamefont
  {O.}~\bibnamefont {Warschkow}}, \bibinfo {author} {\bibfnamefont
  {L.}~\bibnamefont {C~L~Hollenberg}}, \bibinfo {author} {\bibfnamefont
  {G.}~\bibnamefont {Klimeck}}, \ and\ \bibinfo {author} {\bibfnamefont
  {M.}~\bibnamefont {Simmons}},\ }\href {\doibase 10.1038/nnano.2012.21}
  {\bibfield  {journal} {\bibinfo  {journal} {Nat. Nanotechnol.}\ }\textbf
  {\bibinfo {volume} {7}},\ \bibinfo {pages} {242} (\bibinfo {year}
  {2012})}\BibitemShut {NoStop}%
\bibitem [{\citenamefont {Kiczynski}\ \emph {et~al.}(2022)\citenamefont
  {Kiczynski}, \citenamefont {Gorman}, \citenamefont {Geng}, \citenamefont
  {Donnelly}, \citenamefont {Chung}, \citenamefont {He}, \citenamefont
  {Keizer},\ and\ \citenamefont {Simmons}}]{2022Kiczynski}%
  \BibitemOpen
  \bibfield  {author} {\bibinfo {author} {\bibfnamefont {M.}~\bibnamefont
  {Kiczynski}}, \bibinfo {author} {\bibfnamefont {S.~K.}\ \bibnamefont
  {Gorman}}, \bibinfo {author} {\bibfnamefont {H.}~\bibnamefont {Geng}},
  \bibinfo {author} {\bibfnamefont {M.~B.}\ \bibnamefont {Donnelly}}, \bibinfo
  {author} {\bibfnamefont {Y.}~\bibnamefont {Chung}}, \bibinfo {author}
  {\bibfnamefont {Y.}~\bibnamefont {He}}, \bibinfo {author} {\bibfnamefont
  {J.~G.}\ \bibnamefont {Keizer}}, \ and\ \bibinfo {author} {\bibfnamefont
  {M.~Y.}\ \bibnamefont {Simmons}},\ }\href {\doibase
  10.1038/s41586-022-04706-0} {\bibfield  {journal} {\bibinfo  {journal}
  {Nature}\ }\textbf {\bibinfo {volume} {606}},\ \bibinfo {pages} {694}
  (\bibinfo {year} {2022})}\BibitemShut {NoStop}%
\bibitem [{\citenamefont {Wang}\ \emph {et~al.}(2022)\citenamefont {Wang},
  \citenamefont {Khatami}, \citenamefont {Fei}, \citenamefont {Wyrick},
  \citenamefont {Namboodiri}, \citenamefont {Kashid}, \citenamefont {Rigosi},
  \citenamefont {Bryant},\ and\ \citenamefont {Silver}}]{2022Wang}%
  \BibitemOpen
  \bibfield  {author} {\bibinfo {author} {\bibfnamefont {X.}~\bibnamefont
  {Wang}}, \bibinfo {author} {\bibfnamefont {E.}~\bibnamefont {Khatami}},
  \bibinfo {author} {\bibfnamefont {F.}~\bibnamefont {Fei}}, \bibinfo {author}
  {\bibfnamefont {J.}~\bibnamefont {Wyrick}}, \bibinfo {author} {\bibfnamefont
  {P.}~\bibnamefont {Namboodiri}}, \bibinfo {author} {\bibfnamefont
  {R.}~\bibnamefont {Kashid}}, \bibinfo {author} {\bibfnamefont {A.~F.}\
  \bibnamefont {Rigosi}}, \bibinfo {author} {\bibfnamefont {G.}~\bibnamefont
  {Bryant}}, \ and\ \bibinfo {author} {\bibfnamefont {R.}~\bibnamefont
  {Silver}},\ }\href {\doibase 10.1038/s41467-022-34220-w} {\bibfield
  {journal} {\bibinfo  {journal} {Nat. Commun.}\ }\textbf {\bibinfo {volume}
  {13}},\ \bibinfo {pages} {6824} (\bibinfo {year} {2022})}\BibitemShut
  {NoStop}%
\bibitem [{\citenamefont {Le}\ \emph {et~al.}(2020)\citenamefont {Le},
  \citenamefont {Fisher}, \citenamefont {Curson},\ and\ \citenamefont
  {Ginossar}}]{2020Le}%
  \BibitemOpen
  \bibfield  {author} {\bibinfo {author} {\bibfnamefont {N.~H.}\ \bibnamefont
  {Le}}, \bibinfo {author} {\bibfnamefont {A.~J.}\ \bibnamefont {Fisher}},
  \bibinfo {author} {\bibfnamefont {N.~J.}\ \bibnamefont {Curson}}, \ and\
  \bibinfo {author} {\bibfnamefont {E.}~\bibnamefont {Ginossar}},\ }\href
  {\doibase 110.1038/s41534-020-0253-9} {\bibfield  {journal} {\bibinfo
  {journal} {npj Quantum Inf}\ }\textbf {\bibinfo {volume} {6}},\ \bibinfo
  {pages} {24} (\bibinfo {year} {2020})}\BibitemShut {NoStop}%
\bibitem [{\citenamefont {Schofield}\ \emph {et~al.}(2003)\citenamefont
  {Schofield}, \citenamefont {Curson}, \citenamefont {Simmons}, \citenamefont
  {Rue\ss{}}, \citenamefont {Hallam}, \citenamefont {Oberbeck},\ and\
  \citenamefont {Clark}}]{2003Schofield}%
  \BibitemOpen
  \bibfield  {author} {\bibinfo {author} {\bibfnamefont {S.~R.}\ \bibnamefont
  {Schofield}}, \bibinfo {author} {\bibfnamefont {N.~J.}\ \bibnamefont
  {Curson}}, \bibinfo {author} {\bibfnamefont {M.~Y.}\ \bibnamefont {Simmons}},
  \bibinfo {author} {\bibfnamefont {F.~J.}\ \bibnamefont {Rue\ss{}}}, \bibinfo
  {author} {\bibfnamefont {T.}~\bibnamefont {Hallam}}, \bibinfo {author}
  {\bibfnamefont {L.}~\bibnamefont {Oberbeck}}, \ and\ \bibinfo {author}
  {\bibfnamefont {R.~G.}\ \bibnamefont {Clark}},\ }\href {\doibase
  10.1103/PhysRevLett.91.136104} {\bibfield  {journal} {\bibinfo  {journal}
  {Phys. Rev. Lett.}\ }\textbf {\bibinfo {volume} {91}},\ \bibinfo {pages}
  {136104} (\bibinfo {year} {2003})}\BibitemShut {NoStop}%
\bibitem [{\citenamefont {M{\o}ller}\ \emph {et~al.}(2017)\citenamefont
  {M{\o}ller}, \citenamefont {Jarvis}, \citenamefont {Gu\'{e}rinet},
  \citenamefont {Sharp}, \citenamefont {Woolley}, \citenamefont {Rahe},\ and\
  \citenamefont {Moriarty}}]{2017Moller}%
  \BibitemOpen
  \bibfield  {author} {\bibinfo {author} {\bibfnamefont {M.}~\bibnamefont
  {M{\o}ller}}, \bibinfo {author} {\bibfnamefont {S.~P.}\ \bibnamefont
  {Jarvis}}, \bibinfo {author} {\bibfnamefont {L.}~\bibnamefont
  {Gu\'{e}rinet}}, \bibinfo {author} {\bibfnamefont {P.}~\bibnamefont {Sharp}},
  \bibinfo {author} {\bibfnamefont {R.}~\bibnamefont {Woolley}}, \bibinfo
  {author} {\bibfnamefont {P.}~\bibnamefont {Rahe}}, \ and\ \bibinfo {author}
  {\bibfnamefont {P.}~\bibnamefont {Moriarty}},\ }\href {\doibase
  10.1088/1361-6528/28/7/075302} {\bibfield  {journal} {\bibinfo  {journal}
  {Nanotechnology}\ }\textbf {\bibinfo {volume} {28}},\ \bibinfo {pages}
  {075302} (\bibinfo {year} {2017})}\BibitemShut {NoStop}%
\bibitem [{\citenamefont {Achal}\ \emph {et~al.}(2018)\citenamefont {Achal},
  \citenamefont {Rashidi}, \citenamefont {Croshaw}, \citenamefont {Churchill},
  \citenamefont {Taucer}, \citenamefont {Huff}, \citenamefont {Cloutier},
  \citenamefont {Pitters},\ and\ \citenamefont {Wolkow}}]{2018Achal}%
  \BibitemOpen
  \bibfield  {author} {\bibinfo {author} {\bibfnamefont {R.}~\bibnamefont
  {Achal}}, \bibinfo {author} {\bibfnamefont {M.}~\bibnamefont {Rashidi}},
  \bibinfo {author} {\bibfnamefont {J.}~\bibnamefont {Croshaw}}, \bibinfo
  {author} {\bibfnamefont {D.}~\bibnamefont {Churchill}}, \bibinfo {author}
  {\bibfnamefont {M.}~\bibnamefont {Taucer}}, \bibinfo {author} {\bibfnamefont
  {T.}~\bibnamefont {Huff}}, \bibinfo {author} {\bibfnamefont {M.}~\bibnamefont
  {Cloutier}}, \bibinfo {author} {\bibfnamefont {J.}~\bibnamefont {Pitters}}, \
  and\ \bibinfo {author} {\bibfnamefont {R.~A.}\ \bibnamefont {Wolkow}},\
  }\href {\doibase 10.1038/s41467-018-05171-y} {\bibfield  {journal} {\bibinfo
  {journal} {Nat. Comm.}\ }\textbf {\bibinfo {volume} {9}},\ \bibinfo {pages}
  {2778} (\bibinfo {year} {2018})}\BibitemShut {NoStop}%
\bibitem [{\citenamefont {Randall}\ \emph {et~al.}(2018)\citenamefont
  {Randall}, \citenamefont {Owen}, \citenamefont {Fuchs}, \citenamefont {Lake},
  \citenamefont {Ehr}, \citenamefont {Ballard},\ and\ \citenamefont
  {Henriksen}}]{2018Randall}%
  \BibitemOpen
  \bibfield  {author} {\bibinfo {author} {\bibfnamefont {J.~N.}\ \bibnamefont
  {Randall}}, \bibinfo {author} {\bibfnamefont {J.~H.}\ \bibnamefont {Owen}},
  \bibinfo {author} {\bibfnamefont {E.}~\bibnamefont {Fuchs}}, \bibinfo
  {author} {\bibfnamefont {J.}~\bibnamefont {Lake}}, \bibinfo {author}
  {\bibfnamefont {J.~R.~V.}\ \bibnamefont {Ehr}}, \bibinfo {author}
  {\bibfnamefont {J.}~\bibnamefont {Ballard}}, \ and\ \bibinfo {author}
  {\bibfnamefont {E.}~\bibnamefont {Henriksen}},\ }\href {\doibase
  10.1016/j.mne.2018.11.001} {\bibfield  {journal} {\bibinfo  {journal} {Micro
  and Nano Engineering}\ }\textbf {\bibinfo {volume} {1}},\ \bibinfo {pages}
  {1} (\bibinfo {year} {2018})}\BibitemShut {NoStop}%
\bibitem [{\citenamefont {Warschkow}\ \emph {et~al.}(2016)\citenamefont
  {Warschkow}, \citenamefont {Curson}, \citenamefont {Schofield}, \citenamefont
  {Marks}, \citenamefont {Wilson}, \citenamefont {Radny}, \citenamefont
  {Smith}, \citenamefont {Reusch}, \citenamefont {McKenzie},\ and\
  \citenamefont {Simmons}}]{2016Warschkow}%
  \BibitemOpen
  \bibfield  {author} {\bibinfo {author} {\bibfnamefont {O.}~\bibnamefont
  {Warschkow}}, \bibinfo {author} {\bibfnamefont {N.~J.}\ \bibnamefont
  {Curson}}, \bibinfo {author} {\bibfnamefont {S.~R.}\ \bibnamefont
  {Schofield}}, \bibinfo {author} {\bibfnamefont {N.~A.}\ \bibnamefont
  {Marks}}, \bibinfo {author} {\bibfnamefont {H.~F.}\ \bibnamefont {Wilson}},
  \bibinfo {author} {\bibfnamefont {M.~W.}\ \bibnamefont {Radny}}, \bibinfo
  {author} {\bibfnamefont {P.~V.}\ \bibnamefont {Smith}}, \bibinfo {author}
  {\bibfnamefont {T.~C.~G.}\ \bibnamefont {Reusch}}, \bibinfo {author}
  {\bibfnamefont {D.~R.}\ \bibnamefont {McKenzie}}, \ and\ \bibinfo {author}
  {\bibfnamefont {M.~Y.}\ \bibnamefont {Simmons}},\ }\href {\doibase
  10.1063/1.4939124} {\bibfield  {journal} {\bibinfo  {journal} {J. Chem.
  Phys.}\ }\textbf {\bibinfo {volume} {144}},\ \bibinfo {pages} {014705}
  (\bibinfo {year} {2016})}\BibitemShut {NoStop}%
\bibitem [{\citenamefont {Wyrick}\ \emph {et~al.}(2022)\citenamefont {Wyrick},
  \citenamefont {Wang}, \citenamefont {Namboodiri}, \citenamefont {Kashid},
  \citenamefont {Fei}, \citenamefont {Fox},\ and\ \citenamefont
  {Silver}}]{2022Wyrick}%
  \BibitemOpen
  \bibfield  {author} {\bibinfo {author} {\bibfnamefont {J.}~\bibnamefont
  {Wyrick}}, \bibinfo {author} {\bibfnamefont {X.}~\bibnamefont {Wang}},
  \bibinfo {author} {\bibfnamefont {P.}~\bibnamefont {Namboodiri}}, \bibinfo
  {author} {\bibfnamefont {R.~V.}\ \bibnamefont {Kashid}}, \bibinfo {author}
  {\bibfnamefont {F.}~\bibnamefont {Fei}}, \bibinfo {author} {\bibfnamefont
  {J.}~\bibnamefont {Fox}}, \ and\ \bibinfo {author} {\bibfnamefont
  {R.}~\bibnamefont {Silver}},\ }\href {\doibase 10.1021/acsnano.2c08162}
  {\bibfield  {journal} {\bibinfo  {journal} {ACS Nano}\ }\textbf {\bibinfo
  {volume} {16}},\ \bibinfo {pages} {19114} (\bibinfo {year}
  {2022})}\BibitemShut {NoStop}%
\bibitem [{\citenamefont {Liu}\ \emph {et~al.}(2016)\citenamefont {Liu},
  \citenamefont {Lei}, \citenamefont {Shao}, \citenamefont {Ming},
  \citenamefont {Xu}, \citenamefont {Wang},\ and\ \citenamefont
  {Xiao}}]{2016Liu}%
  \BibitemOpen
  \bibfield  {author} {\bibinfo {author} {\bibfnamefont {Q.}~\bibnamefont
  {Liu}}, \bibinfo {author} {\bibfnamefont {Y.}~\bibnamefont {Lei}}, \bibinfo
  {author} {\bibfnamefont {X.}~\bibnamefont {Shao}}, \bibinfo {author}
  {\bibfnamefont {F.}~\bibnamefont {Ming}}, \bibinfo {author} {\bibfnamefont
  {H.}~\bibnamefont {Xu}}, \bibinfo {author} {\bibfnamefont {K.}~\bibnamefont
  {Wang}}, \ and\ \bibinfo {author} {\bibfnamefont {X.}~\bibnamefont {Xiao}},\
  }\href {\doibase 10.1088/0957-4484/27/13/135704} {\bibfield  {journal}
  {\bibinfo  {journal} {Nanotechnology}\ }\textbf {\bibinfo {volume} {27}},\
  \bibinfo {pages} {135704} (\bibinfo {year} {2016})}\BibitemShut {NoStop}%
\bibitem [{\citenamefont {Bennett}\ \emph {et~al.}(2009)\citenamefont
  {Bennett}, \citenamefont {Warschkow}, \citenamefont {Marks},\ and\
  \citenamefont {McKenzie}}]{2009Bennett}%
  \BibitemOpen
  \bibfield  {author} {\bibinfo {author} {\bibfnamefont {J.~M.}\ \bibnamefont
  {Bennett}}, \bibinfo {author} {\bibfnamefont {O.}~\bibnamefont {Warschkow}},
  \bibinfo {author} {\bibfnamefont {N.~A.}\ \bibnamefont {Marks}}, \ and\
  \bibinfo {author} {\bibfnamefont {D.~R.}\ \bibnamefont {McKenzie}},\ }\href
  {\doibase 10.1103/PhysRevB.79.165311} {\bibfield  {journal} {\bibinfo
  {journal} {Phys. Rev. B}\ }\textbf {\bibinfo {volume} {79}},\ \bibinfo
  {pages} {165311} (\bibinfo {year} {2009})}\BibitemShut {NoStop}%
\bibitem [{\citenamefont {Pavlova}\ \emph {et~al.}(2022)\citenamefont
  {Pavlova}, \citenamefont {Shevlyuga}, \citenamefont {Andryushechkin},\ and\
  \citenamefont {Eltsov}}]{2022Pavlova}%
  \BibitemOpen
  \bibfield  {author} {\bibinfo {author} {\bibfnamefont {T.}~\bibnamefont
  {Pavlova}}, \bibinfo {author} {\bibfnamefont {V.}~\bibnamefont {Shevlyuga}},
  \bibinfo {author} {\bibfnamefont {B.}~\bibnamefont {Andryushechkin}}, \ and\
  \bibinfo {author} {\bibfnamefont {K.}~\bibnamefont {Eltsov}},\ }\href
  {\doibase 10.1016/j.apsusc.2022.153080} {\bibfield  {journal} {\bibinfo
  {journal} {Appl. Surf. Sci.}\ }\textbf {\bibinfo {volume} {591}},\ \bibinfo
  {pages} {153080} (\bibinfo {year} {2022})}\BibitemShut {NoStop}%
\bibitem [{\citenamefont {Shevlyuga}, \citenamefont {Vorontsova},\ and\
  \citenamefont {Pavlova}(2023)}]{2023Shevlyuga}%
  \BibitemOpen
  \bibfield  {author} {\bibinfo {author} {\bibfnamefont {V.~M.}\ \bibnamefont
  {Shevlyuga}}, \bibinfo {author} {\bibfnamefont {Y.~A.}\ \bibnamefont
  {Vorontsova}}, \ and\ \bibinfo {author} {\bibfnamefont {T.~V.}\ \bibnamefont
  {Pavlova}},\ }\href {\doibase 10.1021/acs.jpcc.3c00421} {\bibfield  {journal}
  {\bibinfo  {journal} {J. Phys. Chem. C}\ }\textbf {\bibinfo {volume} {127}},\
  \bibinfo {pages} {8978} (\bibinfo {year} {2023})}\BibitemShut {NoStop}%
\bibitem [{\citenamefont {Pavlova}\ and\ \citenamefont
  {Shevlyuga}(2024)}]{2024Pavlova}%
  \BibitemOpen
  \bibfield  {author} {\bibinfo {author} {\bibfnamefont {T.~V.}\ \bibnamefont
  {Pavlova}}\ and\ \bibinfo {author} {\bibfnamefont {V.~M.}\ \bibnamefont
  {Shevlyuga}},\ }\href {\doibase 10.1063/5.0185671} {\bibfield  {journal}
  {\bibinfo  {journal} {J. Chem. Phys.}\ }\textbf {\bibinfo {volume} {160}},\
  \bibinfo {pages} {054701} (\bibinfo {year} {2024})}\BibitemShut {NoStop}%
\bibitem [{\citenamefont {Kresse}\ and\ \citenamefont
  {Hafner}(1993)}]{1993Kresse}%
  \BibitemOpen
  \bibfield  {author} {\bibinfo {author} {\bibfnamefont {G.}~\bibnamefont
  {Kresse}}\ and\ \bibinfo {author} {\bibfnamefont {J.}~\bibnamefont
  {Hafner}},\ }\href {\doibase 10.1103/PhysRevB.47.558} {\bibfield  {journal}
  {\bibinfo  {journal} {Phys. Rev. B}\ }\textbf {\bibinfo {volume} {47}},\
  \bibinfo {pages} {558} (\bibinfo {year} {1993})}\BibitemShut {NoStop}%
\bibitem [{\citenamefont {Kresse}\ and\ \citenamefont
  {Furthm\"uller}(1996)}]{1996Kresse}%
  \BibitemOpen
  \bibfield  {author} {\bibinfo {author} {\bibfnamefont {G.}~\bibnamefont
  {Kresse}}\ and\ \bibinfo {author} {\bibfnamefont {J.}~\bibnamefont
  {Furthm\"uller}},\ }\href {\doibase 10.1103/PhysRevB.54.11169} {\bibfield
  {journal} {\bibinfo  {journal} {Phys. Rev. B}\ }\textbf {\bibinfo {volume}
  {54}},\ \bibinfo {pages} {11169} (\bibinfo {year} {1996})}\BibitemShut
  {NoStop}%
\bibitem [{\citenamefont {Kresse}\ and\ \citenamefont
  {Joubert}(1999)}]{1999Kresse}%
  \BibitemOpen
  \bibfield  {author} {\bibinfo {author} {\bibfnamefont {G.}~\bibnamefont
  {Kresse}}\ and\ \bibinfo {author} {\bibfnamefont {D.}~\bibnamefont
  {Joubert}},\ }\href {\doibase 10.1103/PhysRevB.59.1758} {\bibfield  {journal}
  {\bibinfo  {journal} {Phys. Rev. B}\ }\textbf {\bibinfo {volume} {59}},\
  \bibinfo {pages} {1758} (\bibinfo {year} {1999})}\BibitemShut {NoStop}%
\bibitem [{\citenamefont {Bl\"ochl}(1994)}]{1994Blochl}%
  \BibitemOpen
  \bibfield  {author} {\bibinfo {author} {\bibfnamefont {P.~E.}\ \bibnamefont
  {Bl\"ochl}},\ }\href {\doibase 10.1103/PhysRevB.50.17953} {\bibfield
  {journal} {\bibinfo  {journal} {Phys. Rev. B}\ }\textbf {\bibinfo {volume}
  {50}},\ \bibinfo {pages} {17953} (\bibinfo {year} {1994})}\BibitemShut
  {NoStop}%
\bibitem [{\citenamefont {Perdew}, \citenamefont {Burke},\ and\ \citenamefont
  {Ernzerhof}(1996)}]{1996Perdew}%
  \BibitemOpen
  \bibfield  {author} {\bibinfo {author} {\bibfnamefont {J.~P.}\ \bibnamefont
  {Perdew}}, \bibinfo {author} {\bibfnamefont {K.}~\bibnamefont {Burke}}, \
  and\ \bibinfo {author} {\bibfnamefont {M.}~\bibnamefont {Ernzerhof}},\ }\href
  {\doibase 10.1103/PhysRevLett.77.3865} {\bibfield  {journal} {\bibinfo
  {journal} {Phys. Rev. Lett.}\ }\textbf {\bibinfo {volume} {77}},\ \bibinfo
  {pages} {3865} (\bibinfo {year} {1996})}\BibitemShut {NoStop}%
\bibitem [{\citenamefont {{J{\'o}nsson}}, \citenamefont {{Mills}},\ and\
  \citenamefont {{Jacobsen}}(1998)}]{1998NEB}%
  \BibitemOpen
  \bibfield  {author} {\bibinfo {author} {\bibfnamefont {H.}~\bibnamefont
  {{J{\'o}nsson}}}, \bibinfo {author} {\bibfnamefont {G.}~\bibnamefont
  {{Mills}}}, \ and\ \bibinfo {author} {\bibfnamefont {K.~W.}\ \bibnamefont
  {{Jacobsen}}},\ }\enquote {\bibinfo {title} {Nudged elastic band method for
  finding minimum energy paths of transitions},}\ in\ \href {\doibase
  10.1142/9789812839664_0016} {\emph {\bibinfo {booktitle} {Classical and
  quantum dynamics in condensed phase simulations}}},\ \bibinfo {editor}
  {edited by\ \bibinfo {editor} {\bibfnamefont {B.~J.}\ \bibnamefont
  {{Berne}}}, \bibinfo {editor} {\bibfnamefont {G.}~\bibnamefont {{Ciccotti}}},
  \ and\ \bibinfo {editor} {\bibfnamefont {D.~F.}\ \bibnamefont {{Coker}}}}\
  (\bibinfo  {publisher} {World Scientific},\ \bibinfo {year} {1998})\ pp.\
  \bibinfo {pages} {385--404}\BibitemShut {NoStop}%
\bibitem [{\citenamefont {Brocks}, \citenamefont {Kelly},\ and\ \citenamefont
  {Car}(1992)}]{1992Brocks}%
  \BibitemOpen
  \bibfield  {author} {\bibinfo {author} {\bibfnamefont {G.}~\bibnamefont
  {Brocks}}, \bibinfo {author} {\bibfnamefont {P.}~\bibnamefont {Kelly}}, \
  and\ \bibinfo {author} {\bibfnamefont {R.}~\bibnamefont {Car}},\ }\href
  {\doibase 10.1016/0039-6028(92)91362-F} {\bibfield  {journal} {\bibinfo
  {journal} {Surf. Sci.}\ }\textbf {\bibinfo {volume} {269--270}},\ \bibinfo
  {pages} {860} (\bibinfo {year} {1992})}\BibitemShut {NoStop}%
\bibitem [{\citenamefont {Pavlova}\ and\ \citenamefont
  {Eltsov}(2021)}]{2021Pavlova}%
  \BibitemOpen
  \bibfield  {author} {\bibinfo {author} {\bibfnamefont {T.~V.}\ \bibnamefont
  {Pavlova}}\ and\ \bibinfo {author} {\bibfnamefont {K.~N.}\ \bibnamefont
  {Eltsov}},\ }\href {\doibase 10.1088/1361-648x/ac1092} {\bibfield  {journal}
  {\bibinfo  {journal} {J. Phys.: Condens. Matter}\ }\textbf {\bibinfo {volume}
  {33}},\ \bibinfo {pages} {384001} (\bibinfo {year} {2021})}\BibitemShut
  {NoStop}%
\bibitem [{\citenamefont {Pavlova}\ and\ \citenamefont
  {Shevlyuga}(2023)}]{2023JCPPavlova}%
  \BibitemOpen
  \bibfield  {author} {\bibinfo {author} {\bibfnamefont {T.~V.}\ \bibnamefont
  {Pavlova}}\ and\ \bibinfo {author} {\bibfnamefont {V.~M.}\ \bibnamefont
  {Shevlyuga}},\ }\href {\doibase 10.1063/5.0178757} {\bibfield  {journal}
  {\bibinfo  {journal} {J. Chem. Phys.}\ }\textbf {\bibinfo {volume} {159}},\
  \bibinfo {pages} {214701} (\bibinfo {year} {2023})}\BibitemShut {NoStop}%
\bibitem [{\citenamefont {Cui}\ \emph {et~al.}(2012)\citenamefont {Cui},
  \citenamefont {Gao}, \citenamefont {Jin}, \citenamefont {Zhao}, \citenamefont
  {Tan}, \citenamefont {Fu},\ and\ \citenamefont {Bao}}]{2012Cui}%
  \BibitemOpen
  \bibfield  {author} {\bibinfo {author} {\bibfnamefont {Y.}~\bibnamefont
  {Cui}}, \bibinfo {author} {\bibfnamefont {J.}~\bibnamefont {Gao}}, \bibinfo
  {author} {\bibfnamefont {L.}~\bibnamefont {Jin}}, \bibinfo {author}
  {\bibfnamefont {J.}~\bibnamefont {Zhao}}, \bibinfo {author} {\bibfnamefont
  {D.}~\bibnamefont {Tan}}, \bibinfo {author} {\bibfnamefont {Q.}~\bibnamefont
  {Fu}}, \ and\ \bibinfo {author} {\bibfnamefont {X.}~\bibnamefont {Bao}},\
  }\href {\doibase 10.1007/s12274-012-0215-4} {\bibfield  {journal} {\bibinfo
  {journal} {Nano Res.}\ }\textbf {\bibinfo {volume} {5}},\ \bibinfo {pages}
  {352} (\bibinfo {year} {2012})}\BibitemShut {NoStop}%
\end{thebibliography}%

\end{document}